\documentclass[aps,prl,twocolumn,showpacs,superscriptaddress,floatfix]{revtex4}

\usepackage{graphicx}
\usepackage{epsfig}
\usepackage[english]{babel}
\usepackage{amsmath,amssymb}

\newcommand{\be}{\begin{equation}}
\newcommand{\bea}{\begin{eqnarray}}
\newcommand{\eea}{\end{eqnarray}}
\newcommand{\ee}{\end{equation}}

\def\one{\ensuremath{\hbox{$\mathrm I$\kern-.6em$\mathrm 1$}}}

\begin{document}

\title{Scale perturbation in valence bond ground states}

\author{E. \surname{Rico}}
\affiliation{Institut fuer Theoretische Physik, Universitat Innsbruck, Technikerstrasse 25, A-6020 Innsbruck, Austria.}

\begin{abstract}

A simple and efficient method for calculating the ground state for a class of antiferromagnet systems is presented. It combines the valence bond structure of the ground state for this class of systems and real space renormalization group. As an example, the entire Haldane phase for a spin-1 chain is described, from the AKLT model through to the Heisenberg model, ending at the critical WZNW $SU_2(2)$ model. The picture that emerges from this new method leads us to consider the relationship between the hidden topological order that characterizes the Haldane phase and potential applications in quantum communication technology.

\end{abstract}

\maketitle

One of the main problems in condensed matter physics is the formulation of accurate descriptions of the macroscopic properties of a many-body systems. In order to achieve this aim, it is crucial to understand how do the low energy degrees of freedom emerge from the microscopic dynamics. Much of our knowledge about the emergence of these properties comes from the works by Kadanoff\cite{Kadanoff} and Wilson\cite{Wilson} based on the Renormalization Group (RG) ideas (see also\cite{RG}).

In the last decades, special attention has been given to the analysis of low dimensional antiferromagnetic systems where quantum fluctuations play an essential role (see \cite{lowdim} and references therein). In this context a new structure, known as valence bond, has emerged. This structure describes exactly the ground state of the Affleck-Kennedy-Lieb-Tasaki (AKLT) model\cite{AKLT}, constructed from a Heisenberg-like Hamiltonian. Its inherent quantum nature, however, has brought new perspectives to the field of low dimensional antiferromagnetic systems extending beyond the AKLT model.

The purpose of this work is to introduce a new method to characterize the ground states of quantum systems belonging to this family of states. This method takes the valence bond structure as a building block for ground states and combines it with real space RG in a simple way. We will use the quantum phase characterized by the AKLT model, the Haldane phase\cite{Halda}, to illustrate the ideas coming from this method.

To achieve these goals, we proceed in two steps: first, we will redefine the notion of the valence bond to take into account longer correlations in the system. Second, we will make a scale perturbation on the state to improve the correlations at every scale. The analysis of a spin-1 chain with this method leads to a simple picture of the Haldane phase. We will use this picture to describe the non-local properties of this phase\cite{topo,string} in terms of entanglement\cite{vers}.

An essential requirement for any method that aims to describe the ground state of a quantum system is the efficiency in terms of computational resources and time needed to store the information about the state as well as to extract this information. 

Valence bond states (VBS) in one dimensional systems have a very simple and efficient description within the matrix product states (MPS) formalism\cite{Fannes, MPS}. The goal of this formalism is to describe the wave function so that the entire information is contained in a product of local tensors:
\be
|\Psi\rangle = \sum_{\{s_i\}} A[s_1] |s_1\rangle A[s_2] |s_2\rangle \cdots A[s_n] |s_n\rangle \cdots
\ee
where the index $n$ points to a particular site in the 1D lattice, $\{|s_n\rangle \}$ is a basis in the local Hilbert space $\mathcal{H}_n=\mathbb{C}^d$, and $A[s_n]=p(s_n) T^{(s_n)}$ are the tensors describing the amount of correlations in the state with $p(s_n)$ some probability amplitude and $\{T^{(s_n)}\}$ a basis in $\mathcal{M}_{D\times D}$. Several properties have been characterized\cite{Fannes} for this set of states; among them, we have that
\begin{enumerate}
\item This set is a dense convex  subset of translational invariant states.
\item A transfer matrix that allows us to extract any expectation values in an efficient way can be easily defined.
\item If a state is the exact ground state of a system, then, it minimizes the energy locally i.e. the energy density is equal to the lowest eigenvalue of the interaction.
\end{enumerate}
These three properties are the basis of this work. The first one allows us to use this class of states to formulate a variational ansatz; the second allows us to extract information from the system by a simple product of matrices; the third one shows that these states usually describe the infrared limit (defined later) for a given quantum phase.

Another concept related with the third property is the notion of a parent Hamiltonian\cite{AKLT,cond}. This is the Hamiltonian for which a variational wave function is the exact ground state. The parent Hamiltonian can yield some insight into the symmetries and properties of the phase under study; however, extracting the entire spectrum of eigenvectors is usually a hard task and therefore to apply any perturbation method around the parent Hamiltonian is also difficult. With the method we are introducing, we will see that this task can be solved in an efficient way by defining a non-local MPS.
 
The first goal we would like to achieve, is to describe the main properties of a quantum state in the neighbourhood around VBS, i.e., when the parent Hamiltonian is modified with an irrelevant perturbation such that the system remains in the same phase. To accomplish this, we redefine the notion of a MPS to take into account longer correlations that might appear as the Hamiltonian changes. The ansatz that we propose is to write the state as follows: 
\be
|\Phi\rangle= \sum_{\{s_i\}} \cdots A[s_0,s_1]  |s_1\rangle  A[s_1,s_2]  |s_2\rangle \cdots 
\ee
where $A[s_{n-1},s_n] = p(s_{n-1},s_n) T^{(s_n)}$, with a non-local amplitude of probability $p(s_{n-1},s_n)$ and $\{T^{(s_n)}\}$ a basis in $\mathcal{M}_{D\times D}$. 

This redefinition with a non-local MPS does not spoil the properties of the local MPS. In fact, this non-local redefinition reduces to the standard one when the probability amplitudes are strictly local. Also, one can define a transfer matrix $E[s_{n-1},s_n]= A^*[s_{n-1},s_n] \otimes A[s_{n-1},s_n]$, so that the information about the state can still be extracted in an efficient way. Moreover, the non-local amplitude $p(s_{n-1},s_n)$ can be used to apply variational methods. Due to the fact that the correlations are attached to the physical indexes, it is straightforward to introduce symmetry arguments that simplify the problem. The drawback in this construction is that it is limited to systems with short correlations. If we try to describe models with longer correlations by introducing more sites in the non-local MPS, the number of parameters grows in an exponential way and the method breaks down.

Nevertheless, there is a way to avoid this problem: the combination of the VBS description of the wave function with real space RG methods. This is the second result of this work.

The non-local MPS allows us to describe short range correlations. We can then improve the long range correlations in an efficient way, by integrating over the short distance degrees of freedom. This is nothing but a RG transformation. We therefore need to define the effective degrees of freedom and the dynamics at every scale in the transformation. 

The effective degrees of freedom are extracted from the non-local MPS by a coarse graining of a small cluster of contiguous lattice sites. For instance, the Hilbert space spanned by two lattice sites is given by
\be
\begin{split}
\sum_{s_{2i},s_{2i+1}} T^{(s_{2i})} p(s_{2i},s_{2i+1}) T^{(s_{2i+1})} & |s_{2i},s_{2i+1}\rangle \\
&= q(\tilde{s}_i) T^{(\tilde{s}_i)} |\tilde{s}_i \rangle,
\end{split}
\ee
where $\{|\tilde{s}_i\rangle= \sum_{s_{2i},s_{2i+1}} U^{(s_{2i},s_{2i+1})}_{\tilde{s}_i}  |s_{2i},s_{2i+1}\rangle \}$ is a basis in the effective Hilbert space.

The effective Hamiltonian is given by a real space RG method called contractor renormalization (CORE) created by Morningstar and Weinstein\cite{CORE}.  By definition, this method keeps the eigenvalues of the low energy sector and produces an optimal truncation operator from the original Hilbert space to the effective one.

The low energy eigenvalues, $\{\epsilon_n\}$, can be obtained by exact diagonalization of the Hamiltonian for several clusters in the coarse grained lattice. In this work, we will see that it is enough to solve the interaction between two clusters or equivalently to diagonalize a four body Hamiltonian. This step is the hardest computational problem of this method.

We first calculate the eigenvectors, $\{|n\rangle\}$, for the four body problem and the effective Hilbert space for the contiguous clusters. CORE then shows that the optimal truncation operator is obtained by a Gramm-Schmidt orthogonalization of the eigenvectors of the Hamiltonian projected on the effective Hilbert space, starting from the ground state upward. In this way, a basis, $\{|\phi_n\rangle \}$, is built in which the first vector overlaps with the lowest energy eigenvector and those above, the second one with the second lowest and those above and so on, i.e.
\be
|\phi_n\rangle=\sum_{m \ge n} \lambda_m |m\rangle.
\ee
Finally, the effective Hamiltonian is written as follows:
\be
H_{\text{eff}}=\sum_n \epsilon_n |\phi_n\rangle \langle \phi_n |.
\ee

At this new scale, we can then proceed to get the best non-local MPS and obtain the new effective system for the next scale. Usually, two situations can occur after several steps in the renormalization. The Hamiltonian either flows to a point where the interaction can be solved locally and the correlation length in the effective lattice model goes to zero, or the system is self-similar at every scale, the correlation length diverges and the mass gap goes to zero: at this point, the system is said to be at the critical point.

This method combining VBS structure and real space RG transformation can be summarized as follows:
\begin{enumerate}
\item Solve the short distance problem with the non-local MPS.
\item Extract the effective Hilbert space from the MPS.
\item Compute exactly the eigenvalues of the long distance problem (the hardest computational step).
\item Extract the truncation operator (CORE method).
\item Obtain the Hamiltonian for the next scale and re-scale the unit of distance and energy.
\item Iterate the procedure.
\end{enumerate}

To illustrate the performance of this method, we will analyze the Haldane phase for a spin-1 chain. Several characteristics make this phase important on its own. First, the Heisenberg model is contained in this phase; this model is the subject of the Haldane conjecture\cite{Halda} which states that systems with half integer and integer spin have completely different behaviour. The former are gapless systems while the latter generate a mass gap that cannot be explained with perturbative methods, as this is a pure quantum effect. Second, this phase is a magnetically disordered phase; it describes spin liquids with a hidden topological structure\cite{topo} that cannot be represented with the usual two point correlators\cite{string,vers}.

The Hamiltonian that we will study is written as follows:
\be
H=\sum_i \vec{S}_i \vec{S}_{i+1} - \beta \sum_i \left( \vec{S}_i \vec{S}_{i+1} \right)^2,
\ee
where the operators $S^{a}$, with $a \in \{x,y,z\}$, correspond to the spin-1 representation of $SU(2)$ and $\beta$ is a real parameter. Depending on the actual value of $\beta$, the behaviour of the system can change drastically. Three important points are described by this Hamiltonian:
\begin{itemize}
\item $\beta=1$. At this value the system is integrable. It is described by the critical Wess-Zumino-Novikov-Witten (WZNW) model on the $SU(2)$ group at the level $k=2$. This model happens to be equivalent to three massless Majorana fermions\cite{WZNW}. 
\item $\beta=0$. This point describes the Heisenberg model, the subject of the Haldane conjecture.
\item $\beta=-\frac{1}{3}$. This value will be the starting point of our analysis. It is the AKLT\cite{AKLT} model which has a VBS  as the ground state.
\end{itemize}

The AKLT model has all the ingredients that allow us to perform the scale perturbation as previously discussed. Using a basis in the spin-1 representation such that $S^{a}|b\rangle = i \epsilon^{abc} |c\rangle$, with $\{a,b,c\}=\{x,y,z\}$ and $\epsilon^{abc}$ is the Levi-Civita tensor, the ground state of the AKLT model can be described locally as
\be
|AKLT\rangle_i=\sum_{s_i=\{x,y,z\}} \frac{1}{\sqrt{3}} \sigma^{s_i} |s_i\rangle,
\ee
where $\sigma^{s}$ are the usual Pauli matrices. This description corresponds to a MPS where $p(s_i)= \frac{1}{\sqrt{3}}$ and $T^{(s_i)}= \sigma^{s_i}$. Although this model is not the infrared fixed point of this phase, it is exponentially close to it\cite{prev} and contains the long range and topological properties of the Haldane phase.

In the next step, we will modify the AKLT Hamiltonian to obtain the Heisenberg model and introduce the non-local probability amplitudes $p(s_i,s_{i+1})$ to describe the state at every scale. A basic analysis of the symmetries that characterize this phase simplifies the problem. $SU(2)$ and parity invariance reduce the number of parameters in the variational amplitude. In addition to these symmetries, the normalization of the state is also required.

The ground state energy per site and the correlation length will be completely characterized in our analysis for different values of the parameter $\beta \in[-\frac{1}{3},1]$. Both quantities can be efficiently extracted from the MPS formalism. In the first plot, we show the ground state energy for the Heisenberg model $(\beta=0)$ and for the WZNW model $(\beta=1)$. The energy of the first model converges to $E_0 \simeq -1.435522(5)$ and the correlation length converges to $\xi \simeq 3.8(0)$. These values can be contrasted with previous numerical estimations\cite{num} where the energy is around $3 \%$ higher.

\begin{figure}[!ht]
\begin{center}
\resizebox{!}{2.7cm}{\includegraphics{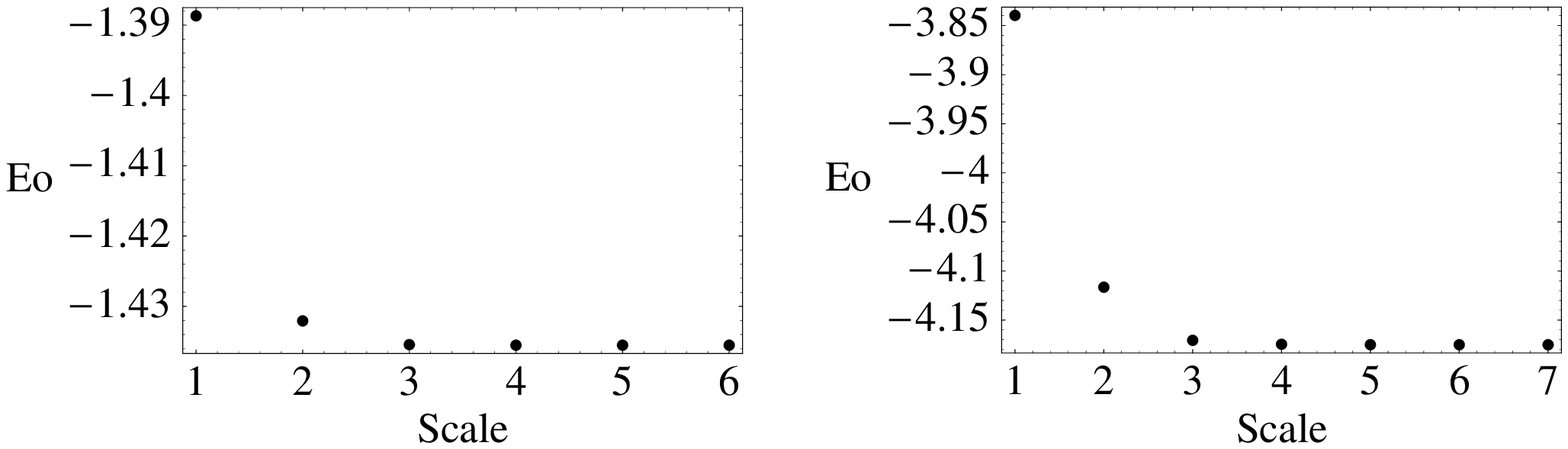}}
\caption{\label{energy} Plots of the ground state energy density as short distance degrees of freedom are integrated. By scale, we mean the number of steps in the renormalization, where we take groups of two contiguous sites every step. The first plot corresponds to the Heisenberg model where the energy per site converges to $E_0 \simeq -1.435522(5)$. The second plot corresponds to the WZNW model where the ground state energy per site converges to $E_0 \simeq -4.175150(1)$.}
\end{center}
\end{figure}

\begin{figure}[!ht]
\begin{center}
\resizebox{!}{2.7cm}{\includegraphics{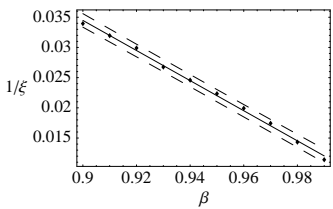}}
\caption{\label{correlator} Plot of the inverse of the correlation length as a function of the coupling constant $\beta$. The points correspond to the numerical data, the solid line to the fitted curve and the dashed lines to a $95\%$ confidence interval for the predicted responses given a linear fit. The inverse of the correlation length goes to zero at the critical value of the coupling constant $\beta_c \simeq 1.03 \pm 0.04$. }
\end{center}
\end{figure}

We can check how good this bound is by comparing the results obtained with another integrable model in the phase, the critical point. The second plot in Fig.\ref{energy} shows that the energy density of the WZNW converges to $E_0 \simeq -4.175150(1)$. In Fig.\ref{correlator}, the behaviour of the inverse of the correlation length is plotted as we approach the critical point $\beta_c \simeq 1.03 \pm 0.04$, at which the correlation length diverges.

\begin{figure}[!ht]
\begin{center}
\resizebox{!}{2.7cm}{\includegraphics{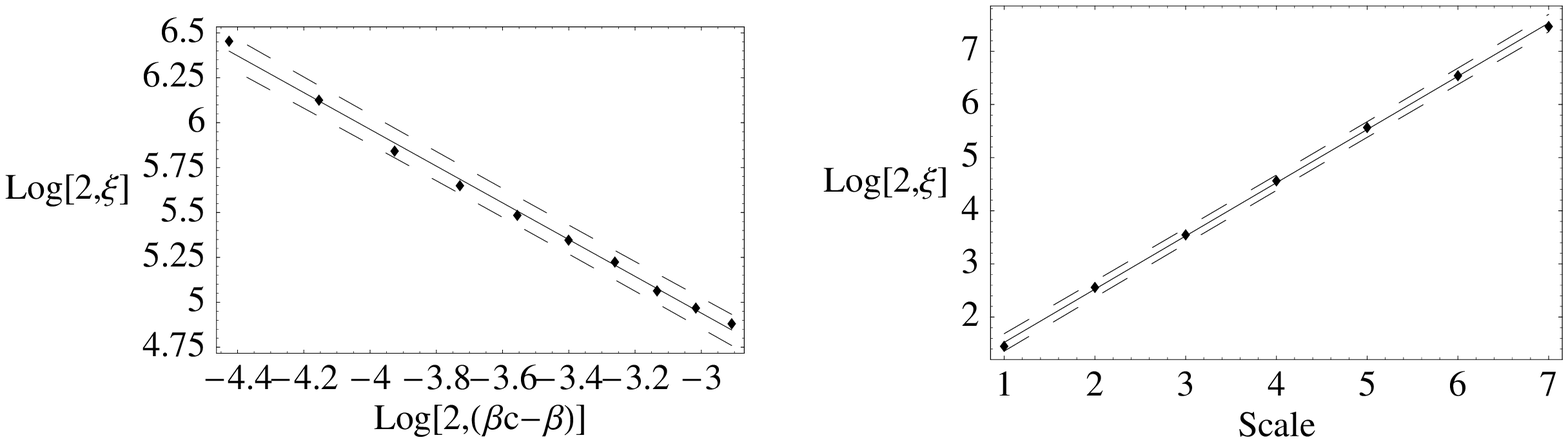}}
\caption{\label{logcorr} Plots of the logarithm of the correlation length in base two. In both cases, the points correspond to the numerical data, the solid line to the fitted curve and the dashed line to a $95\%$ confidence interval for the predicted responses given a linear fit. The first plot shows the logarithm of the correlation length as a function of $\log_2(\beta_c-\beta)$. The slope of the curve corresponds to a critical exponent  $\theta \simeq 1.02 \pm 0.02$. The second plot shows  the logarithm of the correlation length as a function of the scale. The slope of the curve, in this case, corresponds to a critical exponent $\nu \simeq 1.00 \pm 0.01$.}
\end{center}
\end{figure}

Due to the theoretical analysis developed in \cite{WZNW}, we know that the correlation length behaves as $\xi \sim |\beta_c-\beta|^{-\theta}$, with $\theta =1$, as we change the coupling constant close to the critical value $\beta_c$ and $\xi \sim N^{\nu}$, with $\nu=1$, as we modify the infrared cutoff $N=2^{\text{scale}}$. Fig.\ref{logcorr} shows the plots for the behaviour of the correlation length. The slopes in the fitted curves show the scaling of $\xi$ as the coupling constant approaches its critical value with a critical exponent $\theta \simeq 1.02 \pm 0.02$ and the scaling as the infrared cutoff is modified with $\nu \simeq 1.00 \pm 0.01$.

Finally, we would like to comment about the possible consequences of this analysis on the hidden topological order of the Haldane phase\cite{topo} and the potential application of this phase in quantum communication\cite{repeat}. Verstraete and coauthors\cite{vers} showed that the behaviour of the non-local string order parameter of den Nijs and Rommelse\cite{string} can be understood in terms of the special entanglement properties of the Haldane phase. Using the VBS structure of the AKLT model, they showed that it is possible to create a maximally entangled pair with local operations and local measurements at any distance. The work presented here, shows that VBS structure does not change during the renormalization procedure and therefore any ground state in the Haldane phase can be used to obtain a maximally entangled state, at any distance, using local measurement.

To sum up, we have seen how to combine valence bond structure and real space RG. This method allows us to have complete control over the wave function as well as over the dynamics of the system. For that reason, we were able to extract thermodynamical quantities like ground state energy density and correlation length in the entire Haldane phase. Furthermore, we have shown that it is even possible to study critical points in an efficient way.

Recently, three very interesting works\cite{new} about simulation of quantum systems appeared, bringing new insight to the field. These works discuss yet other approaches to this problem.

The author thanks the fruitful discussions at the QIC-UB group in Barcelona and CQC group in Cambridge, where this project started, and at the QIG group in Innsbruck, where this project was finished. Also, the author acknowledges comments and suggestions that appeared in several conversations with Briegel, Cazalilla, Cirac, Ekert, Latorre, Rodriquez and Vidal. Specially, the author thanks the careful reading of the manuscript and valuable remarks by Colbeck, D\"ur and Rodriquez.

This work was supported by the FWF, the European Union (QUPRODIS, OLAQUI, SCALA) and the DFG.


\begin{thebibliography}{99}

\bibitem{Kadanoff} L. Kadanoff. Physics {\bf 2}, 263 (1966); Rev. Mod. Phys. {\bf 39}, 395 (1967).

\bibitem{Wilson} K.G. Wilson. Rev. Mod. Phys. {\bf 47}, 773 (1975).

\bibitem{RG} M.E. Fisher. Rev. Mod. Phys. {\bf 70}, 653 (1998). H.E. Stanley. Rev. Mod. Phys. {\bf 71} S358 (1999).

\bibitem{lowdim} P.W. Anderson. Science {\bf 237}, 1196 (1987). E. Brezin, J. Zinn Justin (ed.) "Fields, Strings and Critical Phenomena" Les Houches, Session XLIX (1989).

\bibitem{AKLT} I. Affleck, T. Kennedy, E.H. Lieb, H. Tasaki. Phys. Rev. Lett. {\bf 59}, 799 (1987); Commun. Math. Phys. {\bf 115}, 477 (1988).

\bibitem{Halda} F.D.M. Haldane. Phys. Lett. {\bf A93}, 464 (1983); Phys. Rev. Lett. {\bf 50}, 1153 (1983).

\bibitem{topo} S.M. Girvin, D.P. Arovas. Physica Scripta {\bf T27}, 156 (1989). 

\bibitem{string} M. den Nijs, K. Rommelse. Phys. Rev. {\bf B40}, 4709 (1989).

\bibitem{vers} F. Verstraete, M. Popp, J.I. Cirac, Phys. Rev. Lett {\bf 92}, 027901 (2004). F. Verstraete, M.A. Martin-Delgado, J.I. Cirac, Phys. Rev. Lett {\bf 92}, 087201 (2004).

\bibitem{Fannes} M. Fannes, B. Nachtergaele, R.F. Werner. Comm. Math. Phys. {\bf 144}, 443 (1992).

\bibitem{MPS} A. Kluemper, A.  Schadschneider, J. Zittartz. EuroPhys. Lett. {\bf 24}, 293 (1993). S. \"Ostlund, S. Rommer. Phys. Rev. Lett. {\bf 75}, 3537 (1995).

\bibitem{cond} A. Auerbach "Interacting electrons and Quantum Magnetism" Springer (1994).

\bibitem{CORE} C.J. Morningstar, M. Weinstein. Phys. Rev. Lett. {\bf 73}, 1873 (1994); Phys. Rev. {\bf D54}, 4131 (1996). 

\bibitem{WZNW} L.A. Takhtajan. Phys. Lett. {\bf A87}, 479 (1982). H. M. Babujan. Nucl. Phys {\bf B215}, 317 (1983). A.M. Tsvelik. Phys. Rev. {\bf B42}, 10499 (1990).   

\bibitem{prev} F. Verstraete et al. Phys. Rev. Lett. {\bf 94}, 140601 (2005). E. Rico. PhD thesis at Barcelona University ({\emph{quant-ph}}/0509037).

\bibitem{num} T. Ziman, H.J. Schulz. Phys. Rev. Lett {\bf 59}, 140 (1987). A. Moreo. Phys. Rev {\bf B35}, 8562 (1987). K. Nomura. Phys. Rev {\bf B40}, 2421(1989). S.R. White. Phys. Rev. {\bf B48} 3844 (1993). M. Weinstein. Nucl. Phys. Proc. Suppl. {\bf 63}, 661 (1998).

\bibitem{repeat} W. D\"ur, H.J. Briegel, J.I. Cirac, P. Zoller. Phys. Rev. {\bf A59}, 169 (1999).

\bibitem{new} G. Vidal {\emph{cond-mat}}/0512165. T.J. Osborne {\emph{quant-ph}}/0601019. F. Verstraete, M.M. Wolf, D. Perez-Garcia, J.I. Cirac {\emph{quant-ph/}}0601075.

\end{thebibliography}
\end{document}